\documentclass[prl,aps,twocolumn,showpacs]{revtex4}
 \newcommand{\mytitle}[1]{
 \twocolumn[\hsize\textwidth\columnwidth\hsize
 \csname@twocolumnfalse\endcsname #1 \vspace{1mm}]}
 \newcommand{\beq}{\begin{equation}}
 \newcommand{\eeq}{\end{equation}}
 \newcommand{\bea}{\begin{eqnarray}}
 \newcommand{\eea}{\end{eqnarray}}

\usepackage{graphics}
\usepackage{graphicx}
\usepackage{amssymb}
\usepackage{bm}

\begin{document}

\title{Nonequilibrium occupation number and charge susceptibility 
of a resonance level close to a dissipative quantum phase transition}
\author{Chung-Hou Chung$^{1,2}$, K.V.P. Latha$^{1,3}$}
\affiliation{
$^{1}$Electrophysics Department, National Chiao-Tung University, HsinChu, Taiwan R.O.C. 300\\
$^{2}$Departments of Physics and Applied Physics, Yale University, New Haven, CT, 06511 USA\\
$^{3}$Institute of applied physics, Academia Sinica, NanKang, Taipei, Taiwan R.O.C. 11529 
}

\date{\today}

\begin{abstract}
Based on the recent paper (Phys. Rev. Lett. {\bf 102}, 216803, (2009)), we 
study the nonequilibrium occupation number $n_d$ and charge susceptibility 
$\chi$ of a resonance level 
close to dissipative quantum phase transition of the Kosterlitz-Thouless (KT) type between a de-localized phase for weak dissipation and a localized 
phase for strong dissipation. 
The resonance level is coupled to two spinless fermionic 
baths with a finite bias voltage and an Ohmic bosonic bath representing the dissipative environment. 
The system is equivalent to an effective anisotropic Kondo 
model out of equilibrium. Within the nonequilibrium Renormalization Group (RG) 
approach, we calculate nonequilibrium 
magnetization $M$ and spin susceptibility $\chi$ 
in the effective Kondo model, corresponding to $2 n_d -1$ and 
$\chi$ of a resonance level, respectively. We demonstrate the smearing 
of the KT transition in the nonequilibrium magnetization $M$ 
as a function of the effective anisotropic Kondo couplings, 
in contrast to a perfect jump in $M$ at the transition in 
equilibrium. In the limit of large bias voltages, we find 
$M$ and $\chi$ at the KT transition and 
in the localized phase 
show deviations from the equilibrium Curie-law behavior. 
As the system gets deeper in the localized phase, both $n_d -1/2$ 
and $\chi$ decrease more rapidly to zero with increasing bias voltages.

\end{abstract}
\maketitle
\subsection {Introduction}

Quantum phase transitions (QPTs)\cite{sachdevQPT,Steve} 
due to competing quantum ground states 
in strongly correlated systems 
have been extensively investigated over the past decades. 
Near the transitions, exotic quantum critical properties are 
realized. In recent years, there has been 
a growing interest in QPTs in nanosystems\cite{lehur1,lehur2,
zarand,Markus,matveev,Zarand2}. 
Very recently, QPTs have been extended to nonequilibrium 
nanosystems where little is known regarding 
nonequilibrium transport near the transitions. A generic 
example\cite{chung} is the transport through a dissipative 
resonance-level (spinless quantum dot)  
at a finite bias voltage where dissipative bosonic 
bath (noise) coming from the environment in the leads gives rise 
to quantum phase transition in transport 
between a conducting (de-localized) phase where resonant tunneling 
dominates and an insulating (localized) phase where 
the dissipation prevails. 
In fact, dissipative quantum phase transitions have been investigated 
in various systems\cite{Josephson,McKenzie}. 
Nevertheless, much of the attention has been 
focused on equilibrium properties; while very little is known 
on the nonequilibrium properties.  
The bias voltage $V$ plays a very different 
role as the temperature $T$ in equilibrium systems as  
the voltage-induced decoherence behaves very differently 
from the decoherence at finite temperature, 
leading to exotic transport properties near the quantum phase 
transition compared to that in equilibrium at finite temperatures.\\

Based on the recent work in Ref. \cite{chung}  
on nonequilibrium transport of a dissipative 
resonance-level at the Kosterlitz-Thouless (KT) type 
de-localized-to-localized quantum transition, 
we study in this paper 
the nonequilibrium occupation number and charge susceptibility 
of a resonance-level quantum dot subjected to a noisy environment near 
the phase transition. In equilibrium, 
it has been shown that the occupation number $n(\epsilon)$ 
of a dissipative resonance-level shows a jump at 
the Fermi energy at the KT transition and in the localized phase 
where $\epsilon$ is an infinitesmall shift 
in the energy of the resonance-level\cite{lehur1, matveev}. 
At finite temperatures 
and in equilibrium, a crossover in $n_d(\epsilon)$ 
replaces the jump and in the high temperature limit it  
is determined by the thermal magnetization of a free spin; 
hence the Curie law behavior is expected. On the other hand, 
when a large bias voltage is 
applied on the system at $T=0$, however, very little is known 
about the nonequilibrium effects on the occupation number 
and charge susceptibility. By first mapping our system onto an 
effective Kondo model and applying the recently developed 
frequency-dependent Renormalization Group (RG) approach\cite{Rosch} 
to the nonequilibrium Kondo effect of a quantum dot, we 
calculate the occupation number and charge susceptibility 
of a resonance-level near the transition. Near the transition, 
we find distinct nonequilibrium 
behaviors of these quantities from those in equilibrium.\\

\subsection {Model Hamiltonian}

The starting point is a spin-polarized quantum dot coupled to two Fermi-liquid 
leads subjected to noisy Ohmic environment, which coupled capacitively to 
the quantum dot. The noisy environment here 
consists of a collection of harmonic oscillators with the Ohmic correlation:
$G_{\phi}({\it i}\omega) \equiv <\phi({\it i}\omega) \phi(-{\it i} \omega)> = 
2\pi \frac{R}{R_k} [|\omega|+\frac{\omega^2}{\omega_c}]^{-1}$ with 
$R$ being the circuit resistance and $R_k\equiv 2\pi \hbar/e^2\approx 25.8k\Omega$ 
being the quantum resistance. 
For a dissipative resonant level (spinless quantum dot) 
model, the quantum phase transition  
separating the conducting and insulating phase for the level is solely 
driven by dissipation. Our Hamiltonian is given by:
\begin{eqnarray}
H &=& \sum_{k,i=1,2} (\epsilon(k)-\mu_i) c^{\dagger}_{k i}c_{k i} + t_{i}
c^{\dagger}_{ki} d + h.c.  \nonumber \\
&+& \sum_{r} \lambda_{r} (d^{\dagger}d-1/2) (b_{r} + b^{\dagger}_{r}) +
\sum_{r} \omega_{r} b^{\dagger}_{r} b_{r},\nonumber\\
&+& h (d^{\dagger} d -1/2)
\end{eqnarray}
where $t_{i}$ is the hopping amplitude between the
lead $i$ and the quantum dot, $c_{ki}$ and $d$ are electron operators for the 
Fermi-liquid leads and the quantum dot, respectively, 
$\mu_i = \pm V/2$ is the chemical potential (bias voltage) 
applied on the lead $i$, while $h$ is the energy level of the dot.
We assume
that the electron spins have been polarized by a 
strong magnetic field. Here, $b_{\alpha}$ are the boson operators of the
dissipative bath with an ohmic spectral density
\cite{lehur2}: $\mathit{J}(\omega) = \sum_{r} \lambda_{r}^2
\delta(\omega-\omega_{r}) = \alpha \omega$ with $\alpha$ being the 
strength of the dissipative boson bath.\\ 
%{\it We use units in which $\hbar=1$ and electric charge $e=1$}.
\begin{figure}[t]
\begin{center}
\includegraphics[width=8.5cm]{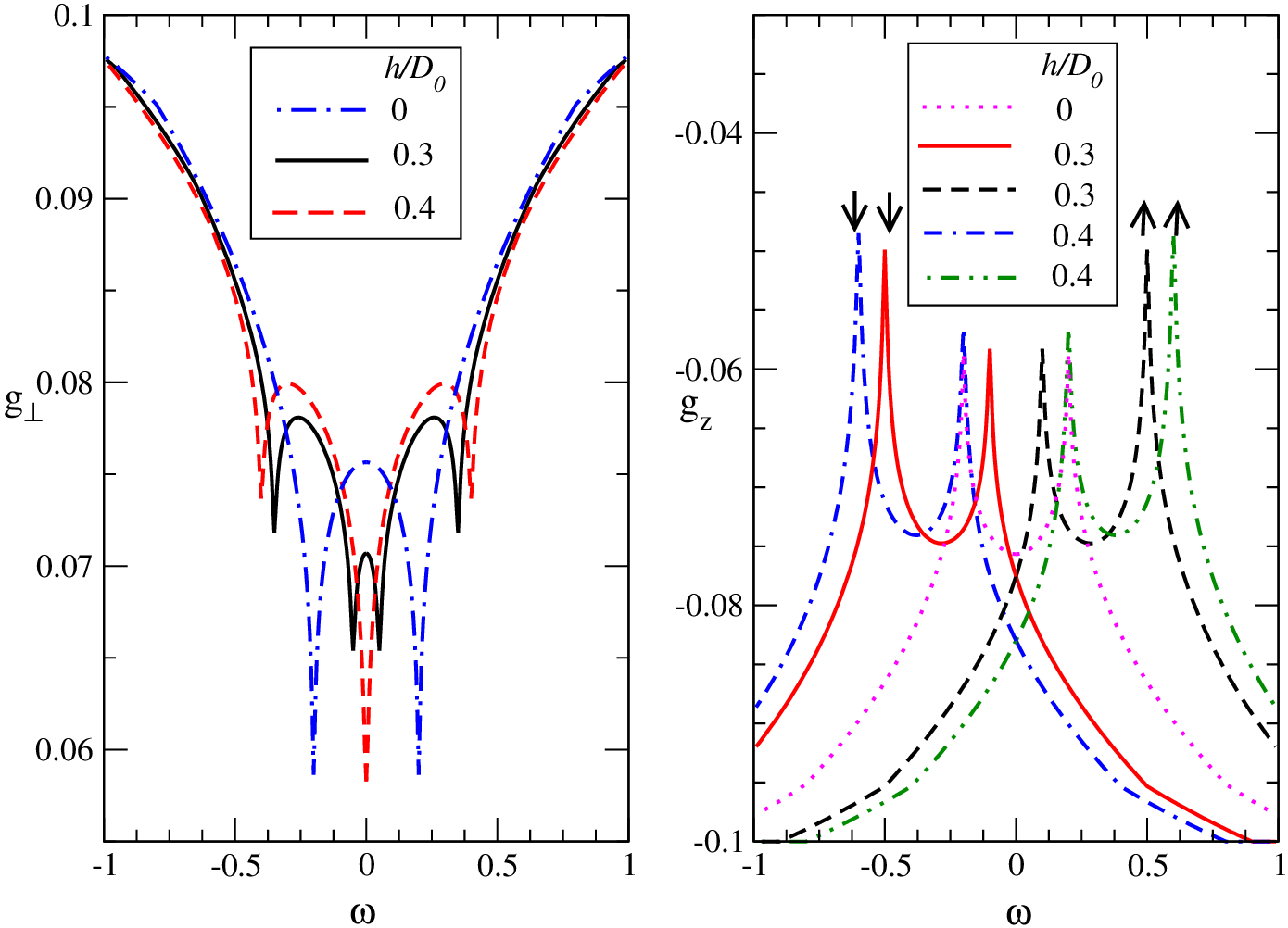}
\end{center}
\par
\vspace{-0.5cm} 
\caption{(Color online) $g_{\perp,z}(\protect\omega)$ versus $\omega$ at the KT transition. 
(a). $g_{\perp}(\omega)$. (b). $g_{z\sigma}(\omega)$. Arrows 
indicate spin $\sigma$ of the corresponding curve.  
The bare couplings are $g_{\perp}= -g_{z} = 0.1 D_0$; bias voltage 
is fixed at $V=0.4 D_0$; the effective magnetic fields are 
fixed at $h=0, 0.3 D_0, 0.4 D_0$.  
Here, $D_0 = 1$ for all the figures.}
\label{gpergzfig}
\end{figure}
First, through similar bosonization and
refermionization procedures as in equilibrium 
\cite{lehur1,lehur2,Markus,matveev}, 
we map our model to an equivalent anisotropic Kondo model in an effective 
magnetic field $h$ with the effective left $L$ and right $R$ Fermi-liquid 
leads\cite{chung}. The effective Kondo 
model takes the form:
\begin{figure}[t]
\begin{center}
\includegraphics[width=8.5cm]{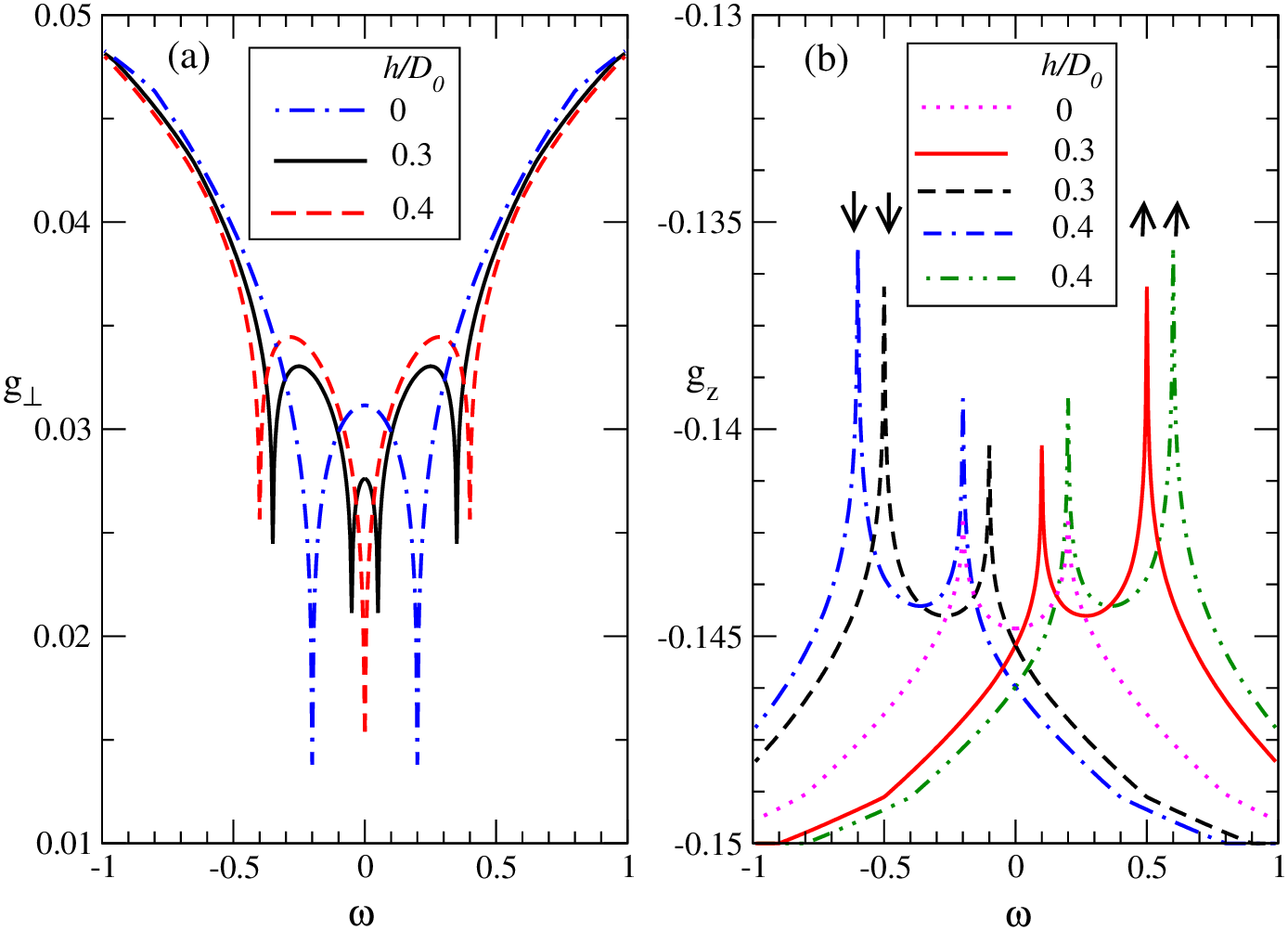}
\end{center}
\par
\vspace{-0.5cm} 
\caption{(Color online) 
$g_{\perp,z}(\protect\omega)$ versus $\omega$ in the localized phase. 
(a). $g_{\perp}(\omega)$. (b). $g_{z\sigma}(\omega)$. Arrows 
indicate spin $\sigma$ of the corresponding curve.  
The bare couplings are $g_{\perp}= 0.05D_0$, $g_{z} = -0.15D_0$; bias voltage 
is fixed at $V=0.4 D_0$; the effective magnetic fields are 
fixed at $h=0, 0.3 D_0, 0.4 D_0$.  
Here, $D_0 = 1$ for all the figures.}
\label{gpergzloc}
\end{figure}
\begin{eqnarray}
{H}_{K} &=&\sum_{k,\gamma =L,R,\sigma =\uparrow ,\downarrow }[\epsilon
_{k}-\mu _{\gamma }]c_{k\gamma \sigma }^{\dagger }c_{k\gamma \sigma } \nonumber \\
&+&(J_{\perp }^{1}s_{LR}^{+}S^{-}+J_{\perp
}^{2}s_{RL}^{+}S^{-}+h.c.)\nonumber \\
&+&\sum_{\gamma =L,R}J_{z}s_{\gamma \gamma
}^{z}S^{z} + h S_z,
\label{Hk}
\end{eqnarray}
where $c_{kL(R)\sigma }^{\dagger }$ is the electron operator of the
effective lead $L(R)$, with spin $\sigma$. 
Here, the spin operators are related to the electron operators on the dot 
by: $S^{+}=d^{\dagger }$, $S^{-}=d$, and $S^{z}=d^{\dagger}d -1/2 = n_d -1/2$ 
where $n_d=d^{\dagger }d$ describes the charge occupancy of the level.
The spin operators for electrons in the effective leads are 
$s_{\gamma \beta }^{\pm }=\sum_{\alpha ,\delta ,k,k^{\prime }}1/2c_{k\gamma
\alpha }^{\dagger }\mathbf{\sigma }_{\alpha \delta }^{\pm }c_{k^{\prime
}\beta \delta }$, 
the transverse and longitudinal Kondo couplings are given by 
$J_{\perp }^{1(2)}\propto {t_{1(2)}}$ and 
$J_{z}\propto 1/2(1-{1}/\sqrt{2\alpha ^{\ast }})$ respectively, 
and the effective bias voltage is 
$\mu _{\gamma }=\pm \frac{V}{2}\sqrt{1/(2\alpha ^{\ast })}$, where $%
1/\alpha ^{\ast }=1+\alpha $. Note that $\mu _{\gamma }\rightarrow \pm V/2$
near the transition ($\alpha ^{\ast }\rightarrow 1/2$ or 
$\alpha \rightarrow 1$) where the above mapping is exact. 
The spin operator of the quantum dot in the effective Kondo model 
$\vec{S}$ can also 
be expressed in terms of spinful pseudofermion operator 
$f_{\sigma}$: 
$S_{i=x,y,z} = f^{\dagger}_{\alpha} \sigma_{i=x,y,z}^{\alpha\beta} f_{\beta}$. 
In the Kondo limit where only the singly occupied fermion states are physically 
relevant, a projection onto the singly occupied states is necessary in the 
pseudofermion representation, which can be achieved by introducing 
the Lagrange multiplier $\lambda$ so that $Q=\sum_{\gamma} f^{\dagger}_{\gamma} f_{\gamma}=1$. An observable $\mathcal{A}$ is defined as\cite{Rosch}:
\begin{equation}
<\mathcal{A}>_{Q=1} =lim_{\lambda\to \infty} 
\frac{<\mathcal{A} Q>_{\lambda}>}{<Q>_{\lambda}}
\end{equation}
%Note that the above mapping 
%can be generalized to a resonance-level coupled to spinless Luttinger 
%liquid leads (see Appendix), which has relevance for  
%edge state tunneling in Quantum Hall states. 
In equilibrium, the above anisotropic Kondo model 
exhibits the Kosterlitz-Thouless transition from a de-localized 
phase with a finite conductance $G\approx \frac{1}{2\pi\hbar}$ 
($e=\hbar=1$) 
for $J_{\perp}+ J_z > 0$ to a localized phase for 
$J_{\perp}+ J_z \le 0$ with vanishing conductance. The nonequilibrium 
transport near the KT transition exhibits distinct profile from that 
in equilibrium and it has been addressed in Ref. \cite{chung}.  
We will focus here on the nonequilibrium occupation number and 
charge susceptibility near the KT transition. 
At the KT transition ($J_{\perp}=-J_z$) and in the localized phase, 
we expect in equilibrium 
a perfect jump in $<n_d>$ (or $<S_z>$): $<n_d> = 1$ for $h>0$ 
and $<n_d>=0$ for $h<0$\cite{lehur1}. At a finite bias voltage, however, 
instead of a jump 
we expect $<n_d>$ shows a smooth crossover as a function of $h/V$. \\

\subsection {Nonequilibrium RG formalism}

\begin{figure}[t]
\begin{center}
\includegraphics[width=8cm]{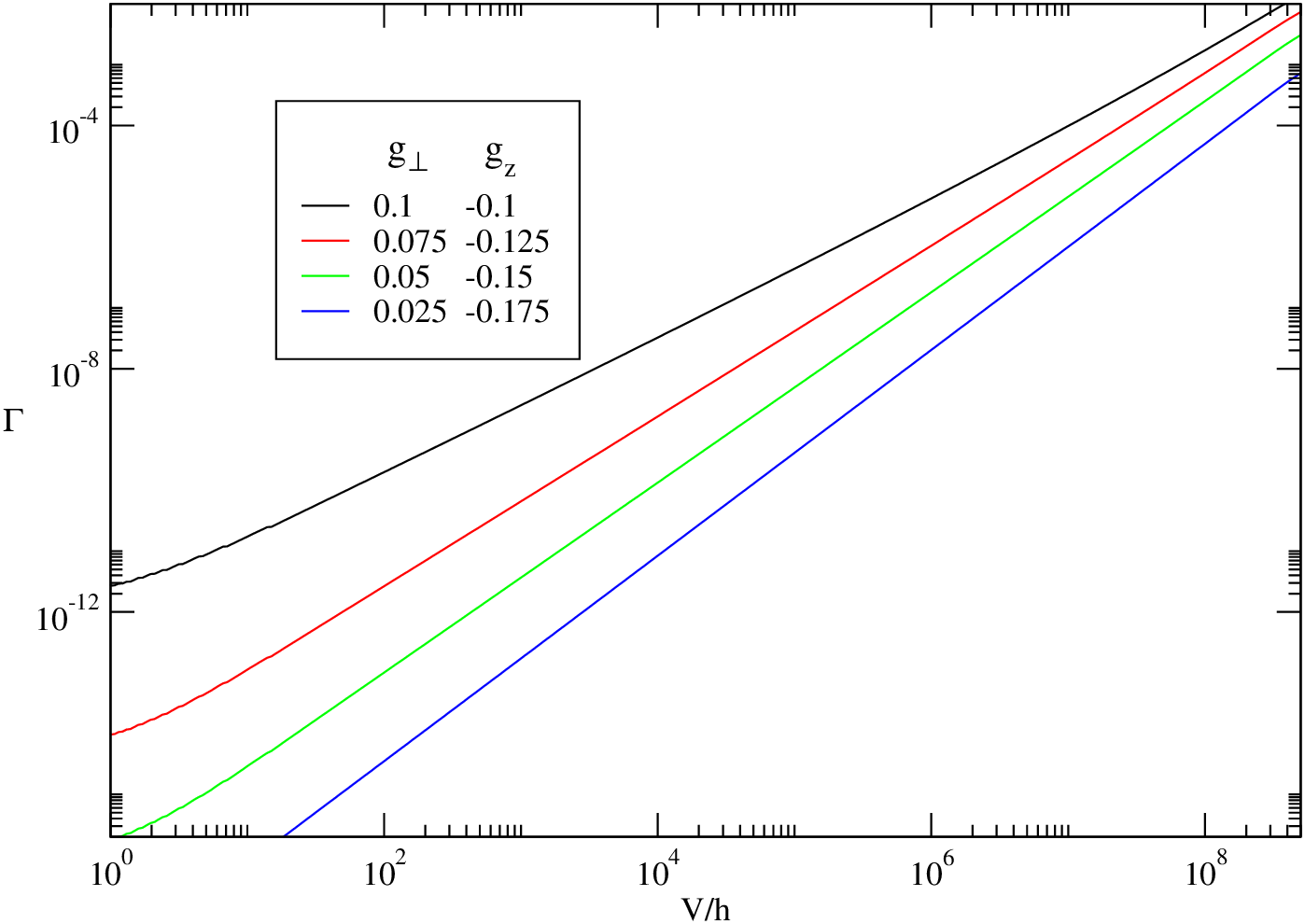}
\end{center}
\par
\vspace{-0.5cm} 
\caption{(Color online) $\Gamma$ versus $V/h$ for different 
bare Kondo couplings. Here, the bare Kondo couplings $g_{\perp,z}$ 
are in units of $D_0$, and $h=10^{-9} D_0$ with $D_0=1$.} 
\label{Gammafig}
\end{figure}
The non-equilibrium perturbative renormalization group (RG) 
equations for the effective Kondo model in a magnetic field 
are obtained by considering the generalized 
frequency dependent Kondo couplings in the Keldysh 
formulation followed Ref.~\cite{Rosch}:

\begin{eqnarray}
\frac{\partial g_{\sigma, z}(\omega )}{\partial \ln D} &=&-\frac{1}{2}\sum_{\sigma\beta =-1,1}\left[
g_{\sigma\perp }\left( \frac{\beta V+\sigma h}{2}\right) \right] ^{2}\Theta _{\omega +
\sigma[h+ \frac{\beta V}{2}]}  \label{gpergz} \nonumber \\
\frac{\partial g_{\sigma, \perp }(\omega )}{\partial \ln D} &=&-\frac{1}{2} \sum_{\sigma\beta=-1,1}  g_{\sigma, \perp }\left( \frac{\beta V+\sigma h}{2}\right)
\times \nonumber\\
&& g_{\sigma, z}\left( \frac{\beta V+\sigma h}{2}\right) \Theta _{\omega +\frac{\beta V+\sigma h}{2}},
\end{eqnarray}
where $g_{\perp\sigma}(\omega) = N(0)J_{\perp\sigma}^{1} = N(0)J_{\perp\sigma}^{2}$, 
$g_{z\sigma}(\omega) = N(0) J_{z\sigma}$ are dimensionless 
frequency-dependent Kondo couplings with $N(0)$ being 
density of states per spin of the 
conduction electrons (we assume symmetric hopping 
$t_1=t_2=t$), $\Theta _{\omega }=\Theta (D-|\omega +\mathit{i}\Gamma |)$, $D<D_{0}$
is the running cutoff, and $\Gamma $ is the decoherence (dephasing) rate at
finite bias which cuts off the RG flow \cite{Rosch}, given by
\begin{eqnarray}
\Gamma &=& 
\sum_{\sigma } 
\frac{\pi}{4\hbar} \int{d\omega} f^{L}_{\omega} \left(1-f^{L}_{\omega} \right)
[g_{\sigma, z}(\omega)]^2 + \nonumber \\
& & f^{L}_{\omega-\sigma h/2} \left(1-
 f^{R}_{\omega+\sigma h/2} \right)
[g_{\perp}(\omega)]^2
+ (L\rightarrow R)
\label{gamma}
\end{eqnarray}
where $f_\omega $ is the Fermi function given by 
$f(\omega) = 1/(1+e^{\omega/kT}))$. Note that the Kondo couplings 
exhibit the following symmetries: $g_{\sigma,\perp}(\omega) = 
g_{-\sigma,\perp}(\omega) = g_{\sigma,\perp}(-\omega)\equiv 
g_{\perp}(\omega)$, $g_{\sigma, z}(\omega) = 
g_{-\sigma, z}(-\omega)$. We have solved the RG equations 
subject to Eq. \ref{gamma} self-consistently. The solutions for 
$g_{\perp}(\omega)$ and $g_{\sigma, z}(\omega)$ at the transition 
are shown in Fig. \ref{gpergzfig}. Similar behaviors for 
$g_{z\sigma,\perp}(\omega)$ are obtained in the localized phase. 
The decoherence rate $\Gamma(V/h)$ is plotted 
in Fig. \ref{Gammafig}.\\
\begin{figure}[t]
\begin{center}
\includegraphics[width=8.0cm]{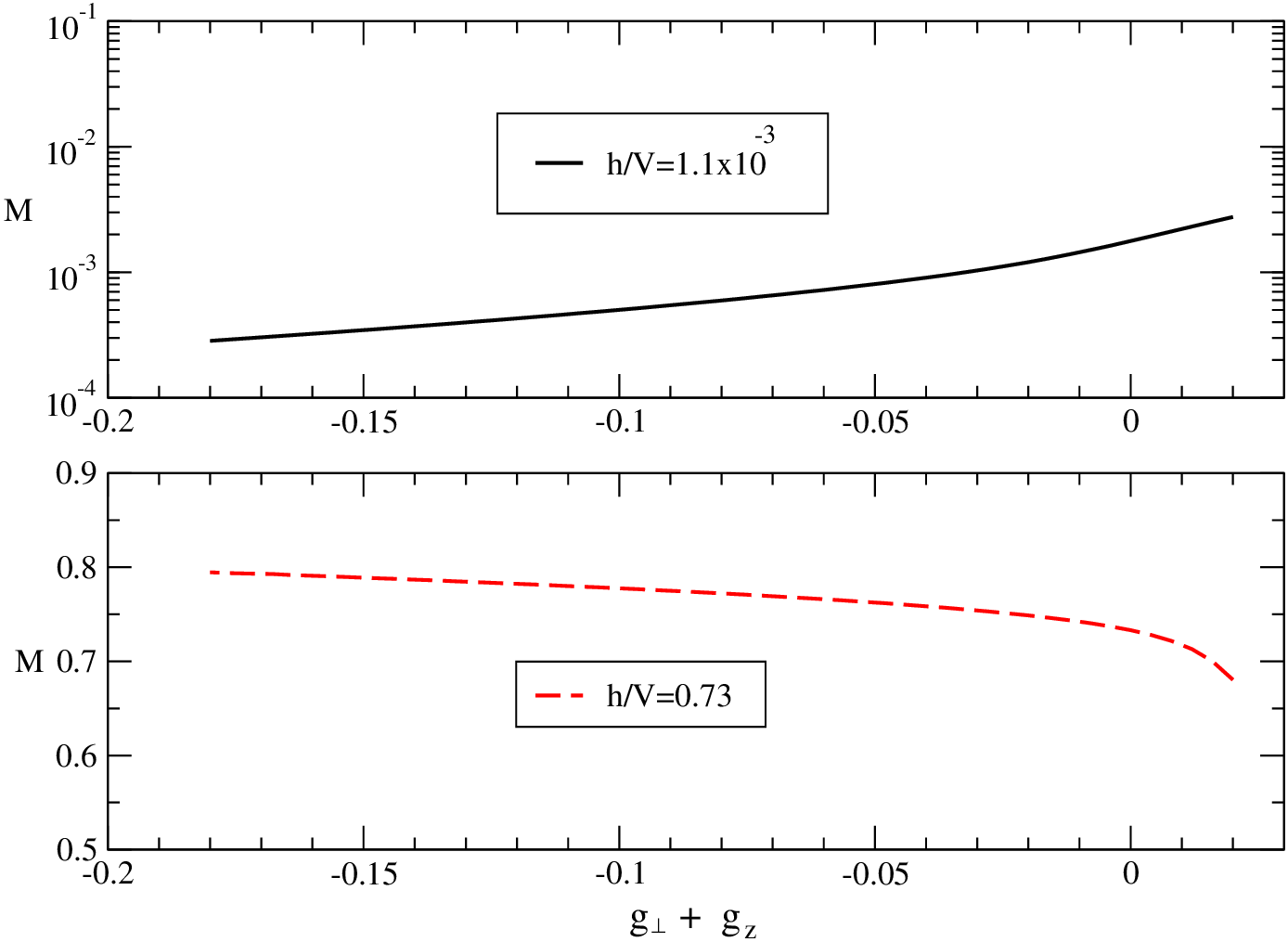}
\end{center}
\par
\vspace{-0.0cm}
\caption{(Color online) 
Nonequilibrium magnetization $M$ in the effective Kondo model 
at fixed $h\approx 9.6\times 10^{-8}D_0$ and fixed bias voltage $V$ 
(small bias with $h/V\approx 0.73$ for lower pannel and large 
bias with $h/V\approx 1.1\times 10^{-3}$ for upper panel) 
versus the initial (bare) Kondo couplings 
$g_{\perp}+g_z$ across the KT transition between the 
delocalized phase ($g_{\perp}+g_z>0$) 
and the localized phase ($g_{\perp}+g_z<0$). 
Here, the bare Kondo couplings $g_{\perp,z}$ 
are in units of $D_0$ with $D_0=1$.}
\label{Mgpgz}
\end{figure}
Note that, unlike the equilibrium RG at finite temperatures 
where RG flows are cutoff by temperature $T$,  
 here in nonequilibrium the RG flows will be cutoff by the 
decoherence rate $\Gamma$, a much lower energy scale than $V$, 
$\Gamma\ll V$. This explains the dips (peaks) structure 
in $g_{\perp (z)}(\omega)$ in Fig. \ref{gpergzfig} and Fig. \ref{gpergzloc}. In contrast, the 
equilibrium RG will lead to approximately frequency independent 
couplings, (or ``flat'' functions 
$g_{\perp}(\omega)\approx g_{\perp,z}(\omega=0)$). 
In the absence of field $h=0$, $g_{\perp,(z)}(\omega)$ show 
dips (peaks) at $\omega=\pm V/2$. 
In the presence of both bias $V$ and field $h$, $g_{\perp}(\omega)$ 
shows dips at $\omega=\pm\frac{V\pm h}{2}$; while 
$g_{z\uparrow(\downarrow)}$ show peaks at $\omega=h\pm V/2$ 
($\omega= -h\pm V/2$)\cite{Rosch}. At $h=V$, two dips of $g_{\perp}(\omega)$ 
at $\omega=\frac{V-h}{2}$ and $\omega=\frac{-V+h}{2}$ merge into 
a large dip at $\omega =0$. 
We use the solutions 
of the frequency-dependent Kondo couplings $g_{\perp, z\sigma}(\omega)$
to compute the occupation number and charge susceptibility 
of the resonance-level near the transition.\\

\subsection {Occupation number and magnetization}

\begin{figure}[t]
\begin{center}
\includegraphics[width=8.0cm]{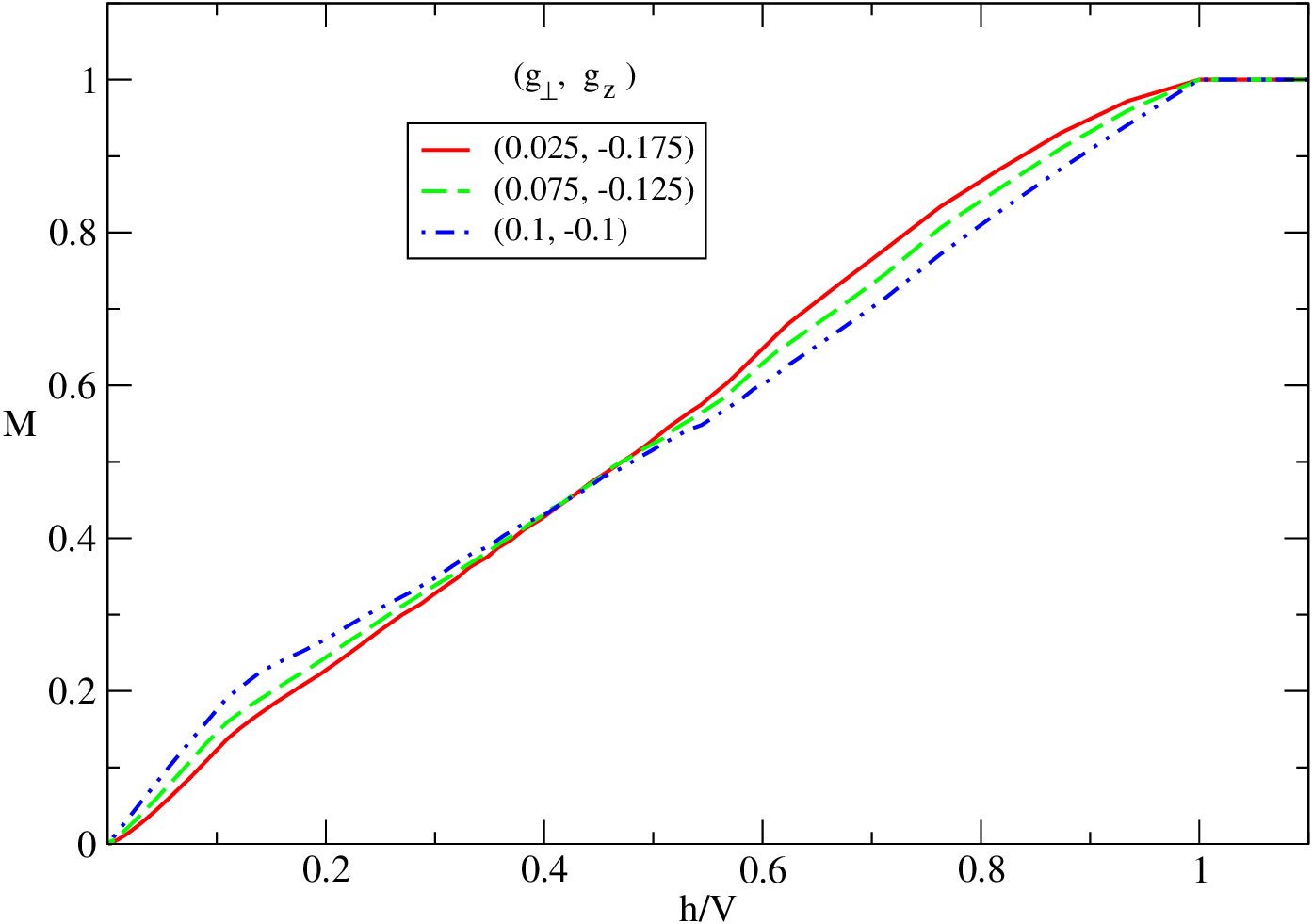}
\end{center}
\par
\vspace{-0.5cm}
\caption{(Color online) 
Nonequilibrium magnetization $M(h/V)$ in 
the effective Kondo model at the KT transition and in the localized 
phase. Here, the bare Kondo couplings $g_{\perp,z}$ 
are in units of $D_0$, and $h=10^{-9} D_0$ with $D_0=1$.}
\label{Mlinear}
\end{figure}
From the mapping, the occupation number of the resonance 
level $n_{d}= d^{\dagger} d$ is related to the magnetization 
of the pseudofermion in the effective Kondo model 
by $S_z = n_d - 1/2 = \frac{M}{2}$ where $M=n_{\uparrow}- n_{\downarrow}= 
f^{\dagger}_{\uparrow} f_{\uparrow} - f^{\dagger}_{\downarrow} f_{\downarrow}$. Since 
the occupation number in the dissipative resonance-level is 
related to the pseudospin magnetization in the effective Kondo model 
by a simple linear relation, in the following we will use the properties 
of the magnetization $M$ to represent 
those of the occupation number. 
The nonequilibrium occupation number of the pseudofermion 
$n_{\uparrow(\downarrow)}=f^{\dagger}_{\uparrow (\downarrow)}f_{\uparrow (\downarrow)}$ 
in the effective model can be determined by solving the Keldysh 
component of the Dyson equation for the pseudofermion 
self-energy\cite{Rosch}, 
given by 
\begin{equation}
n_{\alpha}(\omega) = (1- \Sigma^{>}_{\alpha}(\omega)/\Sigma^{<}_{\alpha}(\omega))^{-1},
\label{Dyson}
\end{equation}
Here, the nonequilibrium pseudofermion self-energies $\Sigma^{<(>)}(\omega)$ 
are obtained via renormalized perturbation theory up to second order 
in $g_{\gamma\gamma'}$:
\begin{eqnarray}
\Sigma_{\alpha}^{<}(\omega) &= &\sum_{\gamma,\gamma' = L,R}{\it i}[n_{\alpha}(-\frac{\alpha h}{2}) 
\chi_{\gamma \gamma'}^{>,z}(-\omega-\frac{\alpha h}{2})\nonumber\\ 
&+&  n_{-\alpha}(\frac{\alpha h}{2}) 
\chi_{\gamma \gamma'}^{>,\perp}(-\omega+\frac{\alpha h}{2})],\nonumber \\
\Sigma_{\alpha}^{>}(\omega) &=& \sum_{\gamma,\gamma' = L,R}-{\it i}[\chi_{\gamma \gamma'}^{<,z}(-\omega-\frac{\alpha h}{2})\nonumber \\
&+&  \chi_{\gamma \gamma'}^{<,\perp}(-\omega+\frac{\alpha h}{2})]
\end{eqnarray}
where
\begin{eqnarray}
\chi_{\gamma\gamma'}^{>,z}(\omega) &=& \int d\epsilon 
[g_{\gamma\gamma'}^z(\epsilon)]^2 \delta_{\gamma\gamma'} f_{\gamma'}(\epsilon) [1-f_{\gamma}(\epsilon+\omega)] \nonumber \\
%\chi_{\gamma\gamma'}^{<,z}(\omega) &=& \int d\epsilon 
%[g_{\gamma\gamma'}^z(\epsilon)]^2 \delta_{\gamma\gamma'}f_{\gamma}(\epsilon+\omega) [1-f_{\gamma'}(\epsilon)]\nonumber \\
\chi_{\gamma\gamma'}^{>,\perp}(\omega) &=& \int d\epsilon 
[g_{\gamma\gamma'}^{\perp}(\epsilon)]^2 \tau^1_{\gamma\gamma'} f_{\gamma'}(\epsilon) [1-f_{\gamma}(\epsilon+\omega)]
%\chi_{\gamma\gamma'}^{<,\perp}(\omega) &=& \int d\epsilon 
%[g_{\gamma\gamma'}^{\perp}(\epsilon)]^2 
%\tau^1_{\gamma\gamma'}f_{\gamma}(\epsilon+\omega) [1-f_{\gamma'}(\epsilon)]
\end{eqnarray}
with $\tau^{1}$ being the $x-$component of the Pauli matrices. 
Similarly, $\chi_{\gamma\gamma'}^{<,z(\perp)}(\omega)$ are obtained 
by interchanging $f_{\gamma'}(\epsilon)$ and $[1-f_{\gamma}(\epsilon+\omega)]$ 
in $\chi_{\gamma\gamma'}^{>,z(\perp)}(\omega)$. 
The nonequilibrium occupation number $n_{\uparrow}(\omega=-\frac{h}{2})$ 
is given by:
\begin{equation}
 n_{\uparrow}(-\frac{h}{2}) = 
\frac{\sum_{\gamma\gamma'} 
\chi_{\gamma\gamma'}^{>,\perp}(h)}
{\sum_{\gamma\gamma'} [\chi_{\gamma\gamma'}^{>,\perp}(h) + 
\chi_{\gamma\gamma'}^{<,\perp}(h)]}
\end{equation}
\begin{figure}[t]
\begin{center}
\includegraphics[width=8.0cm]{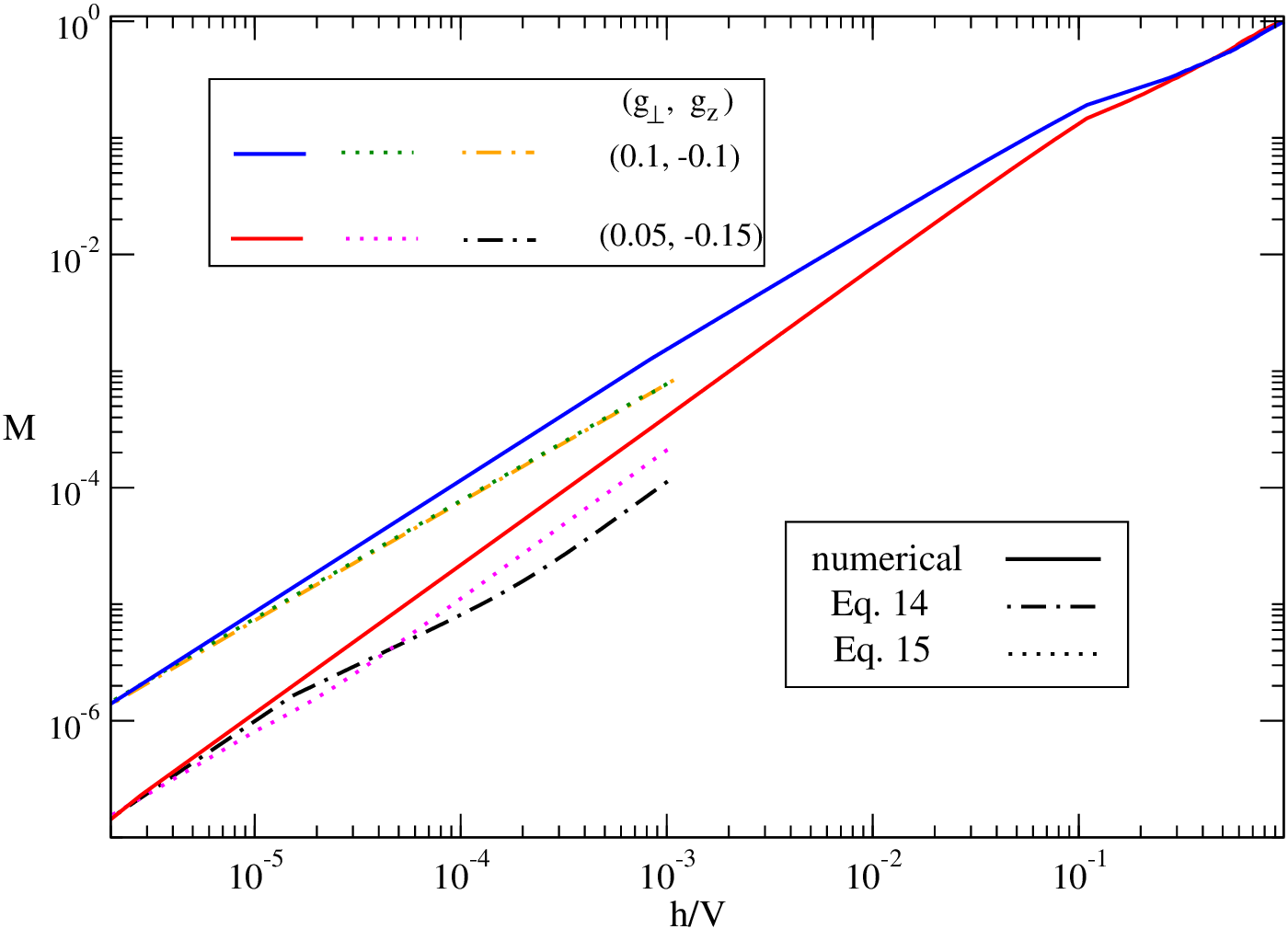}
\end{center}
\par
\vspace{-0.0cm}
\caption{(Color online) 
Nonequilibrium magnetization $M(h/V)$ in 
the effective Kondo model at the KT transition and in the localized 
phase. The dot-dash (dot) lines are results via Eq. 14 (15). 
Here, the bare Kondo couplings $g_{\perp,z}$ 
are in units of $D_0$, and $h=10^{-9} D_0$ with $D_0=1$.}
\label{Mlog}
\end{figure}
The nonequilibrium magnetization $M$ is therefore given by:
\begin{equation}
M  =  
\frac{\sum_{\gamma\gamma'} [\chi_{\gamma\gamma'}^{>,\perp}(h) - 
\chi_{\gamma\gamma'}^{<,\perp}(h)]}
{\sum_{\gamma\gamma'} [\chi_{\gamma\gamma'}^{>,\perp}(h) + 
\chi_{\gamma\gamma'}^{<,\perp}(h)]}
\end{equation}
We can further simplify $M$ as:
\begin{eqnarray}
M &=& \frac{{\mathcal A}- {\mathcal B}}{{\mathcal A}+{\mathcal B}},\nonumber \\
{\mathcal A} &=& \sum_{\alpha \alpha^\prime=L,R}  
\int{d\omega g^2_{\alpha\alpha'\perp}(\omega) f_{\omega-\mu_{\alpha}} \left(
     1 - f_{\omega-\mu_{\alpha^\prime}-\scriptstyle h}\right) } \nonumber \\
{\mathcal B} &=&  \sum_{\alpha \alpha^\prime=L,R} 
    \int{d\omega  g^2_{\alpha\alpha'\perp}(\omega) f_{\omega-\mu_{\alpha}} \left(
     1 - f_{\omega-\mu_{\alpha^\prime}+\scriptstyle h}\right) }\nonumber \\
\label{MT}
\end{eqnarray}
At $T=0$, magnetization $M$ takes the following simple form:
\begin{equation}
M= \frac{\int_{\frac{V-h}{2}}^{\frac{V+h}{2}} d\omega g_{\perp}^2(\omega)
+\int_{\frac{-V-h}{2}}^{\frac{-V+h}{2}} d\omega g_{\perp}^2(\omega) }
{\int_{\frac{-V-h}{2}}^{\frac{V+h}{2}} d\omega g_{\perp}^2(\omega) + 
\int_{\frac{-V+h}{2}}^{\frac{V-h}{2}} d\omega g_{\perp}^2(\omega)}
\label{MT0}
\end{equation}
Note that occupation number  $n_{\uparrow(\downarrow)}$ can also 
be determined by the rate equation\cite{Rosch}: 
$\Gamma_{\uparrow\rightarrow\downarrow} = \Gamma_{\downarrow\rightarrow\uparrow}$ 
or $n_{\uparrow} {\mathcal A} =  n_{\downarrow} {\mathcal B}$ where 
$\Gamma_{\downarrow\rightarrow\uparrow}$ is the spin-flip rate of pseudofermion 
from spin-down to spin-up state. 

We have calculated the magnetization $M(h/V)$ numerically 
at the KT transition and in the localized phase 
for both small bias limit $V\approx h\ll D_0$ and 
large bias limit $V\gg h$ where 
we have fixed $h$ at a small value. 
First, we demonstrate that the nonequilibrium magnetization 
$M$ with a fixed $h\approx 9.6\times 10^{-8}D_0$ for 
both fixed small (lower panel of 
Fig. \ref{Mgpgz}) and large (upper panel of Fig. \ref{Mgpgz}) 
bias voltages shows a smooth crossover as a function of 
$g_{\perp}+g_z$ across the KT transition ($g_{\perp}+g_z=0$), 
in contrast to a perfect jump in $M$ at the transition in 
equilibrium\cite{lehur1}.

\begin{figure}[t]
\begin{center}
%\vspace{1cm}
%\epsfig{file=gpergz.eps,width=0.9\linewidth}
\includegraphics[width=8cm]{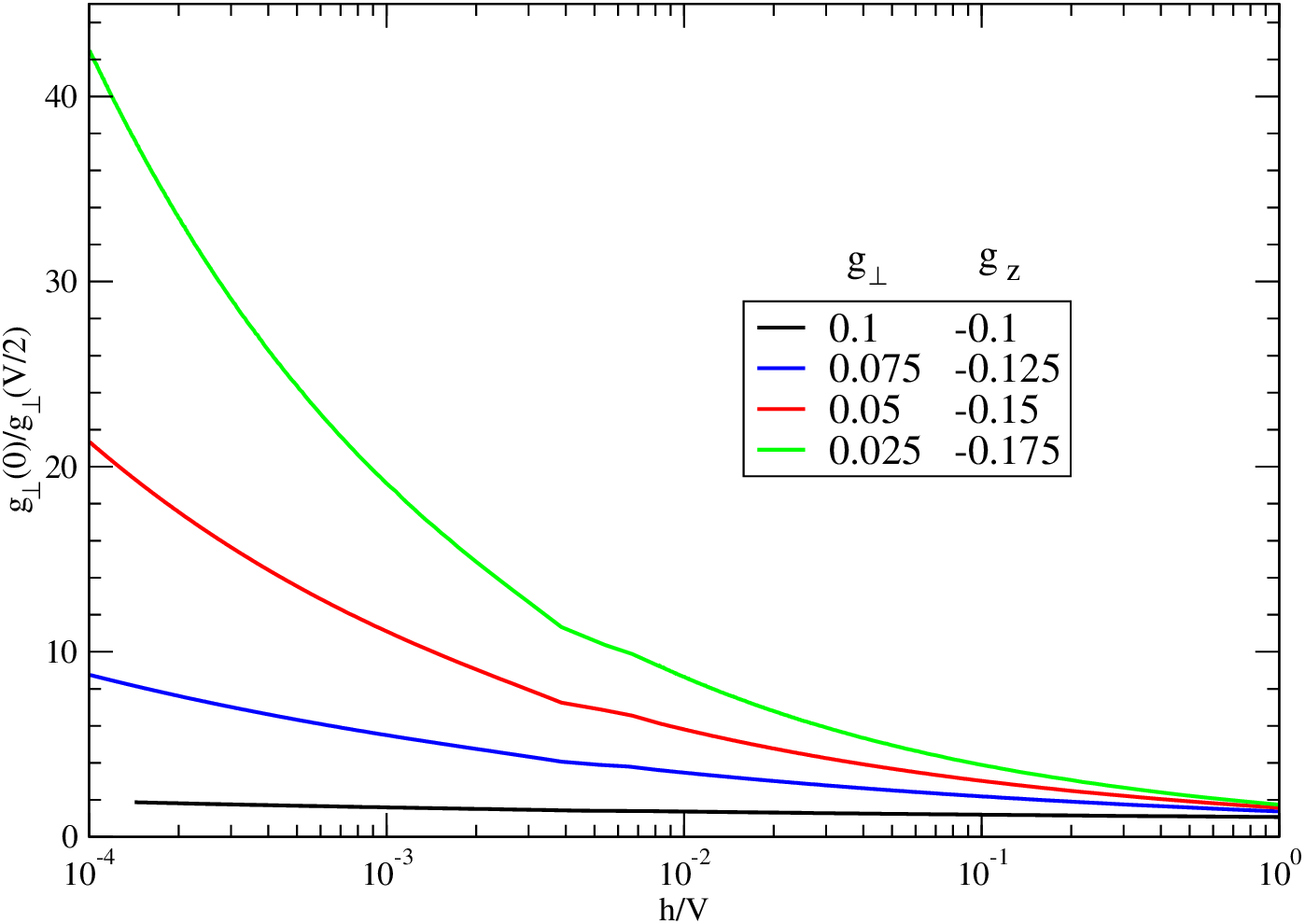}
\end{center}
\par
\vspace{-.0cm} %\label{figgpergz}
\caption{(Color online) $g_{\perp}(\omega=0)/g_{\perp}(\omega=V/2)$ 
versus $h/V$ at the KT transition and in the localized 
phase. Here, the bare Kondo couplings $g_{\perp,z}$ 
are in units of $D_0$, and $h=10^{-9} D_0$ with $D_0=1$.}
\label{gpratio}
\end{figure}

To investigate further the crossover behavior of the magnetization $M$, 
we calculate $M$ as a function of $h/V$ with $h$ being fixed at 
a small value $h\approx 1.0\times 10^{-9} D_0$. 
The result is shown in Fig. \ref{Mlinear}. 
First, let us examine simple limits from the numerical results. 
The spin of the quantum dot gets fully polarized $M=1$ only when magnetic field 
$h$ exceeds the bias voltage, $h\ge V$; while for $h<V$ the 
magnetization is reduced due to finite spin-flip decoherence rate. 
In the extreme large bias limit, $V \gg h$, we find $M$ gets further 
suppression. \\
To gain more understanding of the numerical results, 
we obtain an analytic approximated form for $M(h/V)$ 
for $h\le V$. For $V\rightarrow h$,  
$M\rightarrow 1$ in the following approximated form:
\begin{equation}
M\approx \frac{h}{(V-h)\frac{g_{\perp}^2(0)}{g_{\perp}^2(V/2)} + h}; 
\label{MsmallV}
\end{equation}
while in the large bias limit, $V/h \gg 1$, we find $M(h/V)$ has the 
following approximated form: 
\begin{equation}
M \approx \frac{h g_{\perp}^2(V/2)}{(V-h) \left[\frac{\pi}{4}
       g_{\perp}^2(0)+\left(1 - \frac{\pi}{4}\right)
       g_{\perp}^2(\frac{V-h}{2})\right] +h g_{\perp}^2(V/2)}
\label{MhV}
\end{equation}
Here, we have treated $g_{\perp}(\omega)^2$ within the interval 
$-\frac{V-h}{2}<\omega<\frac{V-h}{2}$ as a semi-ellipse. 
From Eq. \ref{MsmallV} and Eq. \ref{MhV}, it is clear that the behaviors 
of the magnetization $M(h/V)$ depend sensitively on the dip-peak structure in 
$g_{\perp}(\omega)$, especially on the ratio $g_{\perp}(0)/g_{\perp}(V/2)$, 
and $g_{\perp}(\frac{V-h}{2})/g_{\perp}(V/2)$. In general, the 
analytical approximated forms for $g_{\perp}(\omega)$ at 
$\omega=0, \frac{V}{2}, \frac{V-h}{2}$ are rather complex. Nevertheless, 
the values of $g_{\perp}(\omega)$ at 
these specific values of $\omega$ can be obtained numerically 
(see, for example Fig. \ref{gpratio}).\\
In the extremely large bias limit, $h/V\rightarrow 0$, 
$M$ is well approximated by\cite{chung} 
\begin{eqnarray}
M &\approx& \frac{h g_{\perp}^2(V/2)}{V\left[\frac{\pi}{4}
       g_{\perp}^2(0)+\left(1 - \frac{\pi}{4}\right)
       g_{\perp}^2(\frac{V}{2})\right]}
\label{MV}
\end{eqnarray}
In this limit,  
the explicit voltage dependence of $g_{\perp, cr}(\omega=0,V/2)$ 
at the KT transition are given by\cite{chung}:
\begin{eqnarray}
& &g_{\perp,cr}(\omega = 0) \approx \frac{1}{2 \ln({\mathcal{D}}
/V)},\nonumber \\ 
& &g_{\perp,cr}(\omega= V/2) \approx 1/\ln(\frac{{\mathcal{D}}^2}{\Gamma V}),
\label{gpercr}
\end{eqnarray}

Similarly, $g_{\perp}(\omega=0,V/2)_{loc}$ in the 
localized phase take the following forms:
\begin{eqnarray}
& &g_{\perp,loc}(\omega=0) - g_{\perp} \approx \nonumber \\ 
& &   \frac{A}{2 c} [(\frac{V}{D_0})^{2c}\sqrt{c^2+A^2 (\frac{V}{D_0})^{4c}}- \sqrt{A^2+c^2}]\nonumber \\ 
&+& \frac{B}{c} [ (\frac{V}{2D_0})^c \sqrt{c^2+B^2(\frac{V}{2D_0})^{2c}}\nonumber \\ 
&-& (\frac{V}{D_0})^c \sqrt{c^2+B^2 (\frac{V}{D_0})^{2c}}],\nonumber \\
%&+& \frac{c}{2} \ln\left [\frac{[B(\frac{V}{2})^c +
%\sqrt{c^2+B^2 (\frac{V}{2})^{2c}}]}{[BV^{c} +\sqrt{c^2+B^2 V^{2c}}]}\right ] 
%\nonumber \\
%&+& \frac{c}{2} \ln\left [\frac{[A V^{2c} +\sqrt{c^2+
%A^2 V^{4c}}]}{[A +\sqrt{c^2+A^2}]}  \right ]\nonumber \\
\label{gperloc1}
\end{eqnarray}
\begin{eqnarray}
& &g_{\perp,loc}(\omega=V/2) - g_{\perp}\approx \nonumber \\
& &  \frac{A}{2 c} [(\frac{V}{D_0})^{2c}\sqrt{c^2+A^2 (\frac{V}{D_0})^{4c}}- \sqrt{A^2+c^2}]\nonumber \\
 &+& \frac{B}{2c} [(\frac{\Gamma}{D_0})^c \sqrt{c^2+B^2(\frac{\Gamma}{D_0})^{2c}}\nonumber\\
&-& (\frac{V}{D_0})^c \sqrt{c^2+B^2(\frac{V}{D_0})^{2c}}] \nonumber \\ 
%&+& \frac{c}{2} \ln\left[ \frac{(B\Gamma^c +\sqrt{c^2+
%B^2\Gamma^{2c}}) (A V^{2c} +\sqrt{c^2+
%A^2 V^{4c}} ) }{(BV^{c} +\sqrt{c^2+B^2 V^{2c}}) (A +\sqrt{c^2+A^2} ) }
%\right ]\nonumber \\ 
\label{gperloc2}
\end{eqnarray}
where ${\mathcal D} = e^{1/(2 g_{\perp})}$, 
$A = \frac{g_{\perp}}{2} + \frac{cg_{\perp}}{c+|g_z|}$, $B = A V^c$ with $c=\sqrt{g_{z}^2 - g_{\perp}^2}$. Here, we have neglected the subleading terms in 
Eq. \ref{gperloc1} and Eq. \ref{gperloc2} which depend logarithmically 
on $V/D_0$.\\
%Since $\Gamma\ll V$, 
%$g_{\perp}(\omega=V/2) < g_{\perp}(\omega=0)$ and 
%$g_{\perp}(\omega=\frac{V-h}{2}) < g_{\perp}(\omega=0)$ 
%both at the KT transition 
%and in the localized phase, therefore the dips of 
%$g_{\perp}(\omega=\pm \frac{V\pm h}{2})$ 
%lead to further suppession of magnetization from the Curie law behavior. 
%From Eq. \ref{gpercr} and Eq. \ref{gperloc}, the suppression in $M$ 
%at large bias is logrithmic at the KT transition; while 
%it is a more severe power-law fashion in 
%the localized phase as the ratio 
%$\frac{g_{\perp}(\omega=V/2)}{g_{\perp}(\omega=0)}$\cite{chung}: 
We first look at the behavior of $M(h/V)$ at the KT transition. 
At a general level one might expect 
the nonequilibrium magnetization $M(h/V)$ 
at $T=0$ in the de-localized phase 
behave in a similar way as the equilibrium thermal magnetization 
$M(h/T)=\tanh \frac{h}{2T}$ 
with $T$ being replaced by $V$, leading to linear behavior in $h/T$ 
at high temperatures. In equilibrium and at finite temperatures, 
it has been shown that the magnetization of a 
closely related model--a resonance-level with Ohmic dissipation-- 
exhibits linear behavior 
in $h/T$ at the KT transition.  
It is clear from Eq. \ref{MV} that the magnetization $M(h/V)$ in the 
equilibrium form based on the 
"flat approximation" ($g_{\perp}(\omega)\approx g_{\perp}(\omega=0)$) always 
predicts a linear behavior in $h/V$. 
At the KT transition, we find the nonequilibrium 
magnetization $M(h/V)$ for $h\approx V$ 
also shows linear behavior, $M\approx h/V$. 
This can be understood 
from Eq. \ref{MsmallV} as at the KT transition 
$g_{\perp}(0)/g_{\perp}(V/2)\approx 1$ for 
$V\approx h\ll D_0$.  
The Curie-law (linear) behavior in $M(h/V)$ here 
is reminiscent of the equilibrium 
thermal magnetization of a free spin in the high temperature regime. 
However, at the large bias voltages, $V\gg h$, 
we find a logarithmic correction to this linear behavior in $M$ 
at the KT transition  
due to the nonequilibrium 
effect:\\ 
\begin{equation}
M \approx \frac{h}{V} 
\frac{1}{(1-\frac{\pi}{4}) + \frac{\pi}{16} \left(  
\frac{\ln\frac{{\mathcal D}^2}{\Gamma V}}{\ln \frac{{\mathcal D}}{V}} 
\right)^2 }.
\end{equation} 
This logarithmic suppression can be understood 
from Eq. \ref{MV} as in this case $g_{\perp}(\omega=0)$ 
($g_{\perp}(\omega=V/2)$) becomes peak (dip) and 
the ratio satisfies $g_{\perp}(V/2)/g_{\perp}(0) < 1$.

We now discuss $M(h/V)$ in the localized phase.
First, in the limit of small bias, $V\approx h\ll D_0$, as the system 
gets deeper in the localized phase, $M(h/V)$ approaches 
to fully polarization $M=1$ more rapidly than that at the KT transition. 
This is expected 
as the system gets deeper in the localized phase, 
the spin is more easily polarized upon applying a magnetic field.  
This behavior can also be explained from Eq. \ref{MsmallV} 
as in the localized phase the ratio $g_{\perp}(0)/g_{\perp}(V/2)<1$, 
and it only gets smaller as the system gets deeper in the localized phase. 
In fact, the same qualitative behavior is seen in a 
closely related Bose-Fermi Kondo model\cite{lehur1} 
which shows the KT transition between the Kondo and local moment 
ground states. In the large bias limit $V\gg h$, 
however, $M$ deviates from the linear behavior due to nonequilibrium effects. 
 The correction of $M$ to linear behavior is dominated by the ratio 
$g_{\perp}(0)/g_{\perp}(V/2)$ via Eq. \ref{MV} where 
$g_{\perp}(\omega)$ shows deeper dips at $\omega=\pm V/2$, making   
$g_{\perp}(0)/g_{\perp}(V/2)$ to rapidly increase with decreasing 
$h/V$ (see Fig. \ref{gpratio}). 
This gives rise to a further suppression of $M$ at large bias voltages 
compared to that at the KT transition (see Fig. \ref{Mlog}). 
%Note that 
%$M(h/V)$ decreases more slowly in the extreem large bias limit 
%$h/V\rightarrow 0$ as the ratio $g_{\perp}(V/2)/g_{\perp}(0)$ 
%has a maxima as a function of $h/V$ in the large bias region (see Fig. ?).

Note that from Fig. \ref{Mlinear} and Fig. \ref{Mlog}, 
as the system goes deeper into the localized phase 
(or with decrease in $g_{\perp}+g_z <0$),  
we find $M$ for a fixed $h\ll D_0$ 
increases for a fixed small bias voltage ($0.4<h/V<1$); while it decreases 
for a fixed large bias viltage ($h/V\ll 1$). This  
is in perfect agreement with 
the crossover behavior for $M$ shown in Fig. \ref{Mgpgz}.

Notice that the linear behavior of $M(h/V)\approx h/V$ 
is expected in purely asymmetric $g_{LR}> 0 = g_{LL/RR}$ but isotropic 
 ($g_{LL/RR/LR,\perp}=g_{LL/RR/LR,z}$) Kondo model\cite{Rosch}. 
In a symmetric Kondo model with $g=g_{LL}=g_{RR}=g_{LR}$, 
the nonequilibrium magnetization $M(h/V)$ acquires an additional 
positive logarithmic corrections $M\approx (2h/V) (1+ 2 g \ln|V/h|)$
\cite{Rosch}. In the present case, however, the deviation  
from the linear behavior of $M(h/V)$ in the large bias limit 
has a different origin. It comes from the fact that 
our effective Kondo model is not only asymmetric 
($g_{LL/RR}=0<g_{LR}$) but also highly anisotropic 
($g_{LR,z}\le -|g_{LR,\perp}|$) at the KT transition and in the localized phase.
Different corrections to the linear behavior are expected.

\subsection {Susceptibility $\chi$}
 
\begin{figure}[t]
\begin{center}
%\vspace{1cm}
%\epsfig{file=gpergz.eps,width=0.9\linewidth}
\includegraphics[width=8cm]{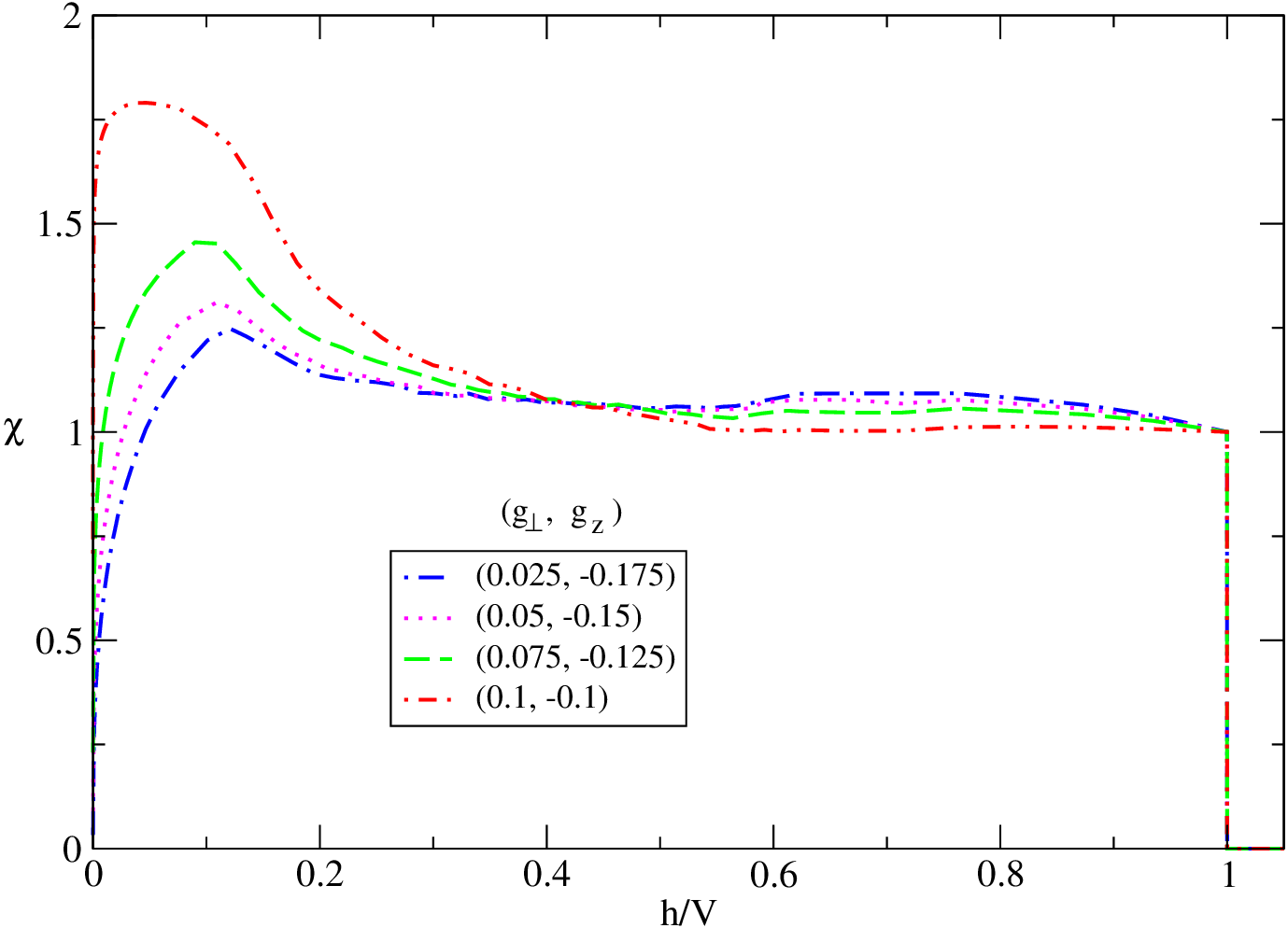}
\end{center}
\par
\vspace{-.0cm} %\label{figgpergz}
\caption{(Color online) $\chi (h/V)$ versus $h/V$ at the 
KT transition and in the localized 
phase. Here, the bare Kondo couplings $g_{\perp,z}$ 
are in units of $D_0$, and $h=10^{-9} D_0$ with $D_0=1$.}
\label{chilinear}
\end{figure}

\begin{figure}[t]
\begin{center}
%\vspace{1cm}
%\epsfig{file=gpergz.eps,width=0.9\linewidth}
\includegraphics[width=8cm]{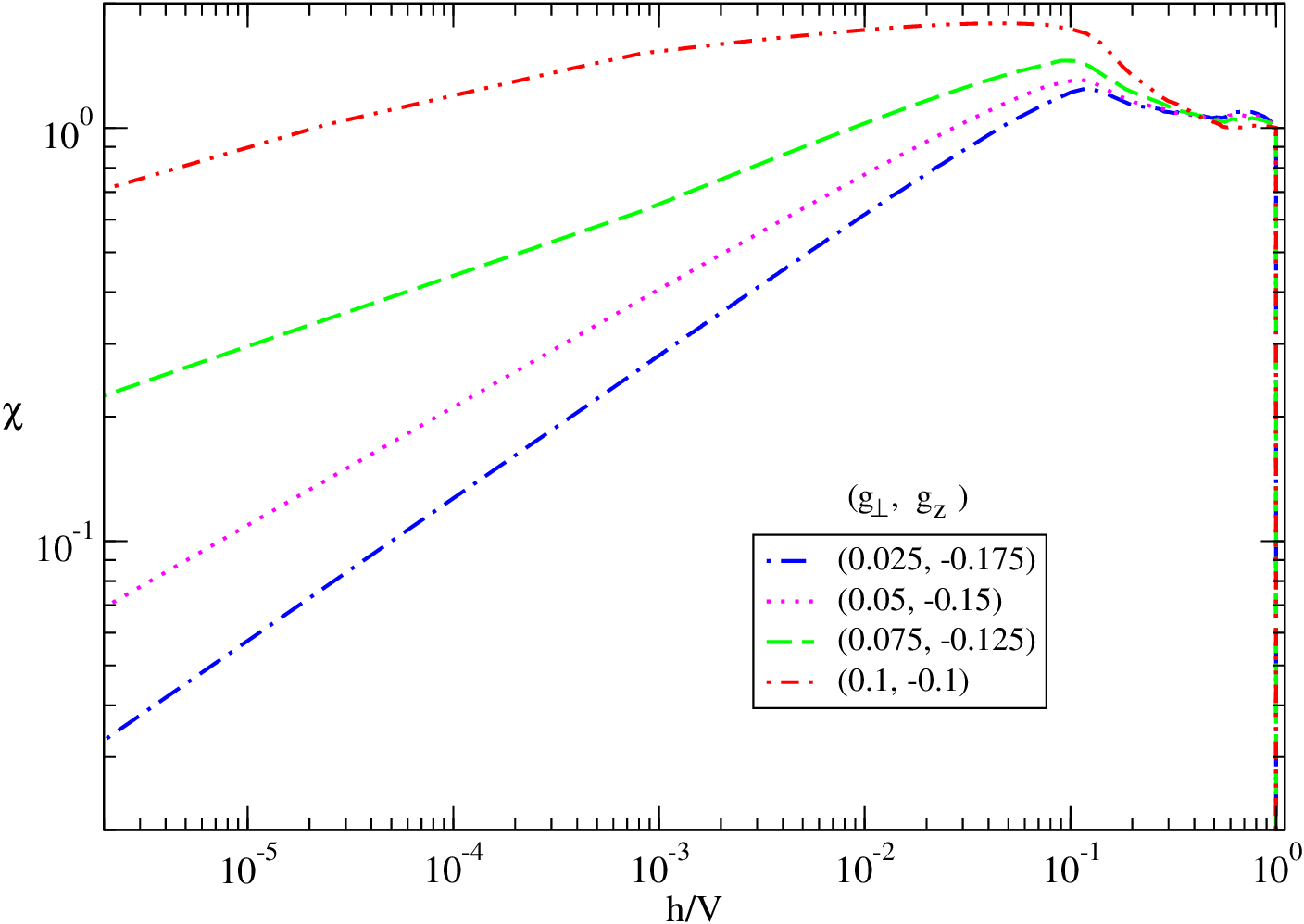}
\end{center}
\par
\vspace{-.0cm} %\label{figgpergz}
\caption{(Color online) $\chi (h/V)$ versus $h/V$ 
at the KT transition and in the localized 
phase. Here, the bare Kondo couplings $g_{\perp,z}$ 
are in units of $D_0$, and $h=10^{-9} D_0$ with $D_0=1$.}
\label{chilog}
\end{figure}

The nonequilibrium charge susceptibility 
$\chi(V)\equiv \frac{\partial n_d }{\partial h}$  
in the dissipative resonance-level 
is obtained from the spin susceptibility   
$\chi = \frac{\partial M}{\partial h}$ in the effective Kondo model by the 
mapping mentioned above. 
The susceptibility $\chi$ of a Kondo dot 
in equilibrium at finite temperatures 
is given by the Curie's law 
$\chi = \frac{1}{2T}$. However, in our highly asymmetric and anisotropic 
Kondo model, we find the nonequilibrium susceptibility 
deviates significantly from the Curie law.
As shown in Fig.\ref{chilinear}, at the KT transition, 
as bias voltage is increased, $\chi(V)$ first shows $1/V$ Curie-law 
behavior, followed by an increase and a peak around $h/V=0.1$. 
In the large bias limit, $\chi(V)$ 
gets a logarithmic suppression (see Eq. \ref{MV}):
\begin{eqnarray}
\chi &\approx& 
\frac{1}{V} \frac{1}{(1-\frac{\pi}{4}) + \frac{\pi}{16} \left(  
\frac{\ln\frac{{\mathcal D}^2}{\Gamma V}}{\ln \frac{{\mathcal D}}{V}} 
\right)^2 } 
\label{chiV}
\end{eqnarray}
Note that the rapid increase in $\chi(h/V)$ at the KT transition 
with decreasing $h/V$ 
for $0.1<h/V<0.5$ is reminiscent 
of the spin susceptibility of a nonequilibrium  
Kondo dot in a magnetic field where $\chi(h/V)$ acquires 
a logarithmic increase at large bias voltages\cite{Rosch}. 
On the other hand, the logarithmic 
decrease in $\chi(h/V)$ here at large bias is a direct consequence 
of the dip structure in $g_{\perp}(\omega)$ at the KT transition 
(see Eq. \ref{gpercr} and Fig. \ref{chilog}).

 As the system gets deeper in the localized phase, 
$\chi(h/V)$ gets a more pronounced peak at  
$h\approx 0.7 V$.  
As bias is further increased, $\chi(h/V)$ shows a similar trend as that 
at the KT transition--a peak around $h/V=0.1$ but smaller magnitudes 
(see Fig. \ref{chilinear}). At large bias voltages, $V\gg h$, 
$\chi$ gets a more severe power-law suppression compared to 
the slower logarithmic decrease at the KT transition (see Fig. \ref{chilog} and 
Eq. \ref{gperloc1} and Eq. \ref{gperloc2}):
\begin{eqnarray}
\chi &\approx& 
\frac{1}{V} \frac{1}{(1-\frac{\pi}{4}) + \frac{\pi}{4}  
\frac{g_{\perp,loc}^2(0)}{g_{\perp,loc}^2(V/2)} } 
\label{chiV}
\end{eqnarray}
 with $g_{\perp,loc}(0), g_{\perp,loc}(V/2)$ given by Eq. \ref{gperloc1} 
and Eq. \ref{gperloc2}.
This comes as a result of 
further decrease in  Kondo coupling  
$g_{\perp}(\omega)$ at $\omega=\pm V/2$ in the localized phase 
under RG.     

We may compare the behavior in $\chi(V)$ in our model 
at large bias voltages with 
that in different limit of the same model or with different models.
In the equilibrium limit 
$V\rightarrow 0$ within our model 
where $g_{\perp,z}(\omega)$ can be considered as 
flat functions over $-\frac{V+h}{2}<\omega<\frac{V+h}{2}$, 
$g_{\perp}(\omega=0)/g_{\perp}(\omega=V/2)\approx 1$, a perfect Curie law behavior is expected 
for $\chi(V)$. However, for isotropic Kondo model ($g_{LL}=g_{RR}=g_{LR}$) for a simple 
quantum dot in Kondo regime and at large bias voltages, $\chi(V)$ shows Curie law with  
positive logarithmic correction, $\chi\approx \frac{1}{V} (1+ \frac{1}{\ln\frac{V}{T_k}})$
with $T_k$ being Kondo temperature for a single quantum dot. In our dissipative 
resonance-level model, the suppression in $\chi(h/V)$ at large bias voltages 
at the KT transition and in the localized phase comes from the dips at 
$g_{\perp}(\omega=\pm V/2)$.

 \begin{center}
{\bf Conclusions}
\end{center}

In conclusion, we have investigated the nonequilibrium 
occupation and charge susceptibility of a dissipative 
resonance-level with energy $h$. For $h=0$, the system 
exhibits the Kosterlitz-Thouless type quantum transition 
between a de-localized phase at small dissipation strength 
and a localized phase with large dissipation. 
We first mapped our problem onto an effective nonequilibrium 
anisotropic Kondo model in the presence of a magnetic field 
$h$. The occupation number and 
charge susceptibility correspond to magnetization $M$ 
and susceptibility $\chi$ 
of the pseudospin in the effective Kondo model, respectively.  
By nonequilibrium RG approach, 
we solved for the frequency-dependent effective 
Kondo couplings and calculated magnetization $M(h/V)$ 
and $\chi(h/V)$ at finite bias voltages. 
We demonstrate the smearing of the KT transition 
in the nonequilibrium 
magnetization $M$ at a fixed $h$ as a function 
of the effective anisotropic Kondo couplings   
for both small bias and large bias voltages as it 
exhibits a smooth crossover at 
the KT transition, in contrast to a perfect jump 
in $M$ at the transition in equilibrium.
For small 
bias $V$ and effective field $h$ and $h\approx V$, 
we find the magnetization $M(h/V)$ 
at the KT transition shows 
linear behavior in $h/V$; while in the localized phase 
$M$ increases more rapidly with $V$ approaching to $h$ from above, 
consistent with the behaviors of the equilibrium 
magnetization in the localized phase at finite temperatures. 
In the large bias limit $V\gg h$, however, we find corrections  
to equilibrium Curie-law behavior in $M$ due to nonequilibrium effects. 
At the KT transition, the corrections are logarithmic; in $V/\cal{D}$; 
while in the localized phase they are power-law in $V/D_0$. Our 
results have direct relevance for the transport measurements 
in nanostructures, and should stimulate further experiments. \\

\acknowledgements

We are grateful for the helpful discussions with 
P. W\"{o}elfle. This work is supported by the NSC grant
No.98-2918-I-009-06, No.98-2112-M-009-010-MY3, the MOE-ATU program, the
NCTS of Taiwan, R.O.C. (C.H.C.).

%\bibitem{Florens}
%Serge Florens, Pascal Simon, Sabine Andergassen, and Denis Feinberg, 
%Phys. Rev. B {\bf 75}, 155321 (2007).

%\bibitem{Lee}
%Yu-Wen Lee and Yu-Li Lee, Phys. Rev. B {\bf 65}, 155324 (2002).

%\bibitem{Ng}
%Yu-Liang LIu and T. K. Ng, Phys. Rev. B {\bf 61}, 2911 (2000).

%\bibitem{Kim}
%Eugene Kim, cond-mat/0106575 (unpublished).

%\bibitem{wen}
%Xiao-Gang WEn, cond-mat/9812431 (unpublished).

\end{document}